\documentclass[12pt]{article}
\usepackage{adjustbox}
\usepackage[toc,page]{appendix}
\usepackage[english]{babel}

\usepackage[letterpaper,top=2cm,bottom=2cm,left=3cm,right=3cm,marginparwidth=1.75cm]{geometry}

\usepackage{amsmath}
\usepackage{graphicx}
\usepackage{caption}
\usepackage{subcaption}
\usepackage[colorlinks=true, allcolors=blue]{hyperref}

\title{Hey That's Mine!\\ Imperceptible Watermarks are Preserved in Diffusion Generated Outputs}
\author {
    Luke Ditria\textsuperscript{\rm 1},
    Tom Drummond\textsuperscript{\rm 2},
}
\begin{document}
\maketitle

\begin{abstract}
Generative models have seen an explosion in popularity with the release of huge generative Diffusion models like Midjourney and Stable Diffusion to the public. Because of this new ease of access, questions surrounding the automated collection of data and issues regarding content ownership have started to build. In this paper we present new work which aims to provide ways of protecting content when shared to the public. We show that a generative Diffusion model trained on data that has been imperceptibly watermarked will generate new images with these watermarks present. We further show that if a given watermark is correlated with a certain feature of the training data, the generated images will also have this correlation. Using statistical tests we show that we are able to determine whether a model has been trained on marked data, and what data was marked. As a result our system offers a solution to protect intellectual property when sharing content online.
\end{abstract}

\section{Introduction}
Image generation has become a booming area of research, with many large image generation models becoming available to the general public and the commercialisation of such models becoming its own industry \cite{ramesh2022hierarchical, rombach2022high, zhang2023adding, saharia2022photorealistic}. Progress in this area has introduced a number of new issues around the effortless creation of realistic images. Such issues include the proliferation of misinformation and the use of digital content without the creator's permission. With the rise of these ready to use text to image generators, high-quality fake images can be quickly produced on mass.\\
\begin{figure}[t]
\centering
\includegraphics[width=0.95\textwidth]{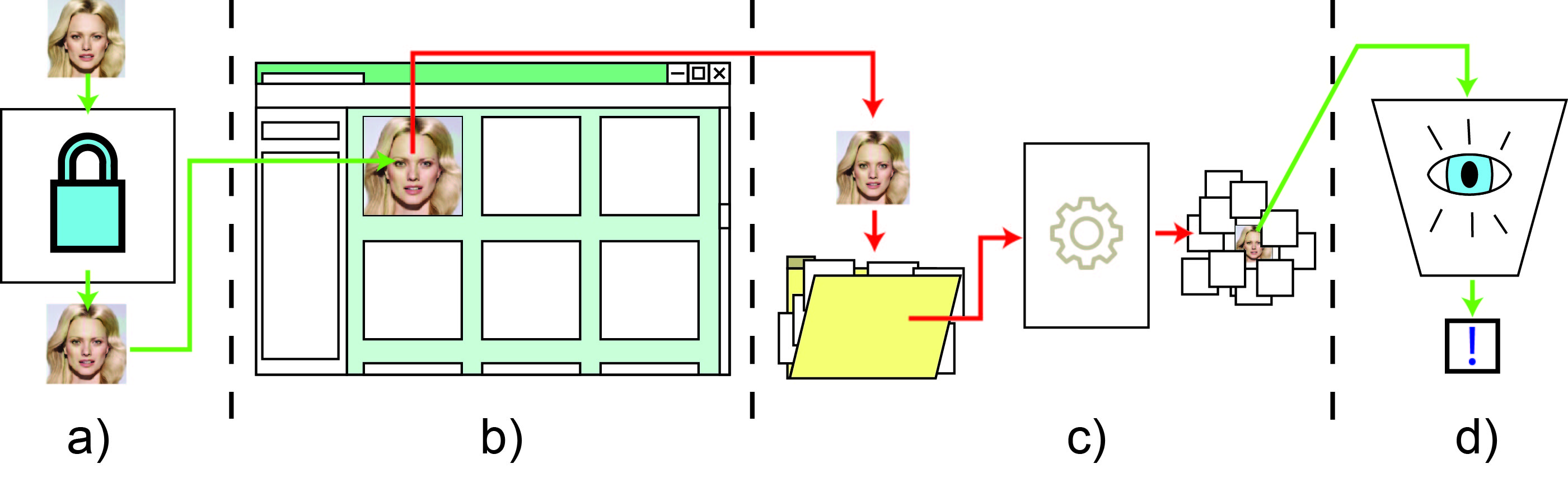}
\caption{Outline of a possible use case: a) Content is encoded with an imperceptible watermark. b) Watermarked image is posted online making it vulnerable to being used without consent. c) Image is taken and used without consent to train a generative model. d) Generated images are decoded to determine if a correlation between features and watermarks exists.}
\label{fig:use_case}
\end{figure}

To help counter this, many investigations been undertaken in order to detect generated images. Such work either focuses on detecting generated images directly with DNN based classification \cite{pu2020noisescope} or marking fake generated images with imperceptible watermarks that a suitable detector can recognise \cite{yu2021artificial, zhao2023recipe, fernandez2023stable}. The first methods rely on there being some perceptible different between generated and real images and struggles when the generated images come from a model other then the one it was trained on. The second method is more robust as the diffusion model itself can be trained to produce images with imperceptible watermarks. However as the watermark is embedded after generation this method relies on the creator of such generative models adding these watermarks as well as making the detector available.\\

The issue of digital content being used without consent when training generative models is a much more difficult problem to solve. Many content creators regularly share their work online, making it easy for individuals or organisations to automatically collect vast quantities of high quality data without consent from the original creators. Even if one visibly watermarks an image, modern Deep learning techniques could automatically detect and remove these. Without access to the data-set that was used to train the generative model, or the parameters of the model itself, it can be very difficult to determine whether or not a given piece of digital content was used. The creators of these generative models could simply claim that any similarity between generated images and a given piece of content is a coincidence. This issue raises major concerns around copyright and intellectual property \cite{henderson2023foundation} and may cause independent creators to become reluctant to share their work, stifling innovation and the sharing of new ideas. \\

As far as we are aware, no work has been undertaken on determining whether images have been created by a generative model that was trained on an individual's content. Such a method would empower creators to identify when their content is being used to train a generative model, without access to the model itself. \\
In this paper we demonstrate that when images are watermarked with a suitable imperceptible watermarking technique, a simple Diffusion Model will generate images with this watermark present without requiring any auxiliary losses. When these watermarks are correlated with a given feature of the data, such as the content or style, the diffusion model will also produce watermarks with the same correlation. This not only allows one to detect whether or not an image is fake, but also allows one to determine which images were used for training. By doing so we hope to open a new field of research with the aim of developing tools to allow better transparency with generative models and allow proper attribution to the original content creators.\\

Specifically in this work we demonstrate:
\begin{itemize}
  \item A watermarking system to imperceptibly mark images that is robust to common data pre-processing methods.
  \item That a diffusion model trained on imperceptibly watermarked images will generate images with these watermarks present.
  \item That when watermarks are correlated with a given feature of the data, the generated images will also contain watermarks with the same correlation.
\end{itemize}
\section{Related work}

\subsection{Diffusion Models}
\begin{figure}
\centering
\includegraphics[width=0.8\textwidth]{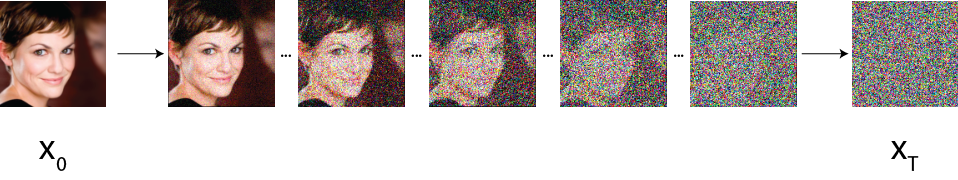}
\caption{Sequentially noising an image in a diffusion forward process.}
\label{fig:Forward_process}
\end{figure}
A recent advancement in the field of generative models is the development of Denoising Diffusion Probabilistic Models (DDPM) \cite{ho2020denoising}. These models have been shown to produce high fidelity results when used on image data and do not suffer from training instabilities like GANs. Diffusion models are characterised by a forward ``Diffusion" process of $T$ steps where Gaussian noise is gradually added to data, i.e an image, until all information has been destroyed and the sample is indistinguishable from a  sample of white noise (Figure \ref{fig:Forward_process}). The model then learns a single step of the ``Reverse Process" at a time, which will remove a small amount of noise. During inference it is then possible to start at a random Gaussian sample and iteratively denoise until we arrive at our generated sample. Simplifications of this method have been proposed such as Denoising Diffusion Implicit Models (DDIM) \cite{song2020denoising}, which convert the stochastic process into a deterministic one. Cold Diffusion \cite{bansal2022cold} improved robustness and demonstrated that it is possible to invert any arbitrary destructive process. Similar to other generative models, Diffusion models can also be conditioned on additional inputs to both improve image quality and control generated outputs \cite{batzolis2021conditional, ho2022classifier}. 
\begin{figure*}[ht]
\centering
\includegraphics[width=0.9\textwidth]{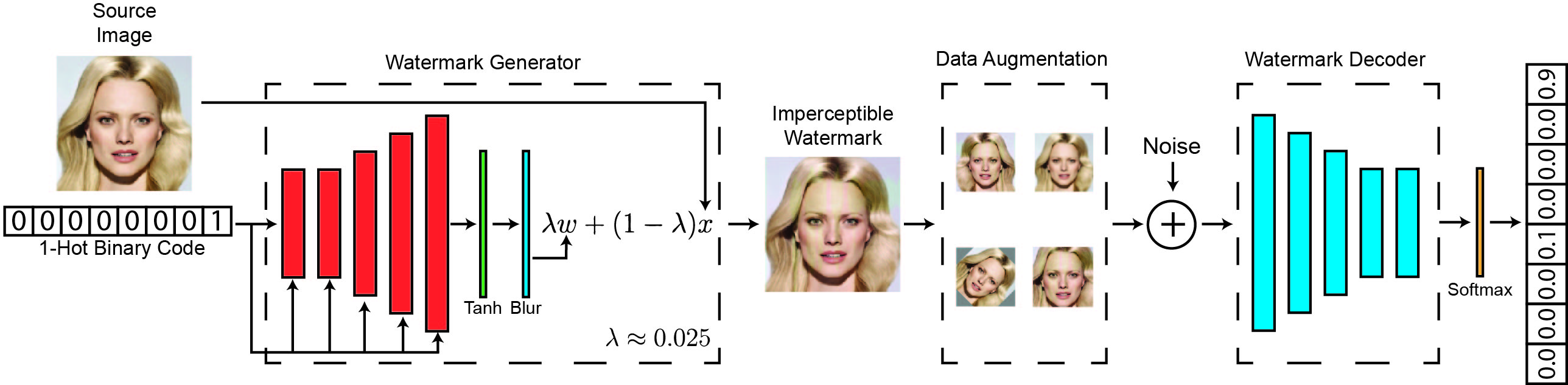}
\caption{Process of training the watermark Generator and Decoder. Random image and watermark pairs are chosen for training the Generator/Decoder system. The watermark Generator creates a unique watermark image for each index and combines it with the target image. To account for common preprocessing steps used when training diffusion models, data augmentation is performed. Both the watermark Generator and Decoder are trained via a Cross-Entropy loss.}
\label{fig:Watermarker}
\end{figure*}
\subsection{Deep Steganography}
Steganography looks at methods of hiding a piece of information within another data source. A typical example of this is hiding a secret message within the pixel values of an image. The information to be hidden is typically referred to as the ``source" and the data in which it is hidden, the ``container". With the rise of Deep learning, many Deep Steganography methods have been developed \cite{chen2023low, baluja2017hiding}. These methods have been able to hide a large amount of information within a single image. Most of these methods jointly train an encoder and decoder network, to both minimise the difference between the encoded container and the original container image as well as the decoder error. Newer methods are able to do away with the container image and instead utilise image generation techniques to directly generate images with the information embedded \cite{wei2023generative, zhang2019steganogan}. 

\subsection{Watermarking To Protect Models}
A large body of work has been undertaken on the related task of DNN watermarking for proving ownership \cite{li2021survey}. Many of these ideas have been applied to Diffusion based image generation models. These methods either augment the training process to directly generate watermarked images for easy fake content detection \cite{wen2023tree, zhao2023recipe}, or aim to hide the watermarking capabilities of the model until a given ``trigger" is provided \cite{liu2023watermarking, peng2023protecting, li2022backdoor} thus proving ownership.
\subsection{Watermarking To Protect Data}
Steganography aims to hide a message in a container, focusing on how to robustly encode as much information as possible. However, some applications do not require a large amount of data to be encoded. If you wish to simply encode a unique watermark onto an image that can be later detected, much simpler techniques can be employed. Furthermore as fewer bits are required, these methods can be made very robust to a variety of augmentation techniques. The problem of determining whether a model has used a specific dataset during training has been explored within the field of DNN based classification \cite{tang2023did, sablayrolles2020radioactive}. Such methods aim to determine whether a model has been trained on a given dataset by the use of watermarking and label manipulation. These methods require manipulation of inputs and targets and are designed to protect a whole dataset.\\
Due to the complexity of the issue the pool of existing work that aims to achieve a similar effect for image generation models is limited. \cite{carlini2023extracting} has shown that it is possible to extract some of the training data from a diffusion model, even when treating the model as a ``black box". However it is not possible to determine from a novel generated image whether or not a given image is present in the hidden dataset. Some methods for watermarking generative models come close to this idea, but focus on marking data to protect the generative model itself and either explicitly train the model to preserve the watermarks \cite{zhao2023recipe} or work on a dataset level and focus on deep fake detection \cite{yu2021artificial}.
\begin{figure}[ht]
\centering
\includegraphics[width=0.8\textwidth]{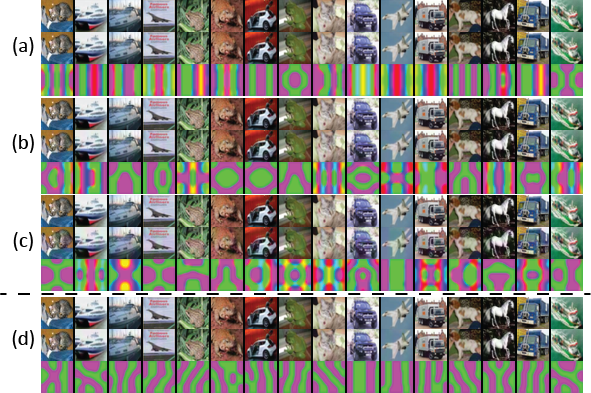}
\caption{Outputs of watermark Generator. For each part; Top: Original, Middle: Watermarked image, Bottom watermark only. a/b/c correspond to watermark scales ($\lambda$) of 0.010/0.025/0.050 respectively, where each model was trained with watermark augmentations. d) was trained without any augmentations after watermarking and $\lambda=0.025$.}
\label{fig:wm_imgs}
\end{figure}
\begin{figure}[ht]
\centering
\includegraphics[width=0.8\textwidth]{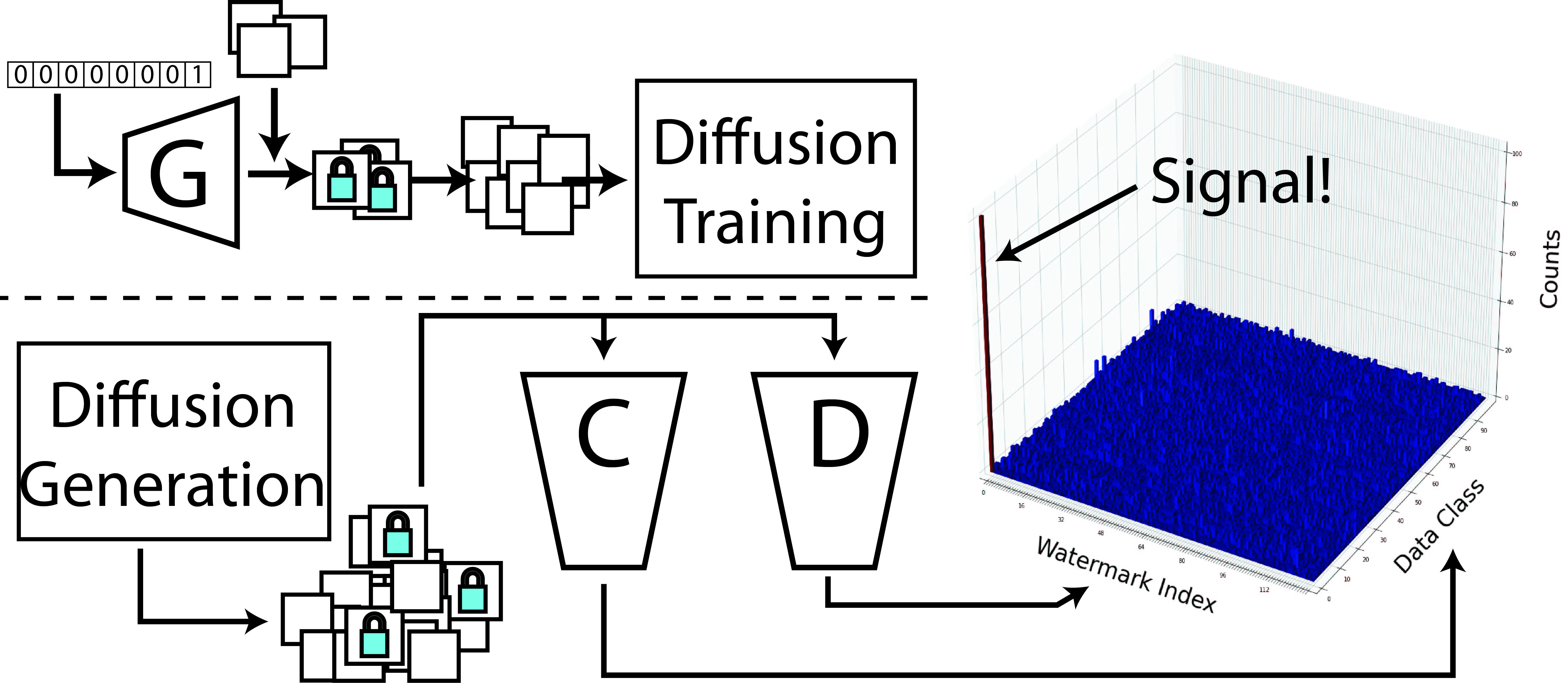}
\caption{Watermarking Diffusion dataset. The watermark Generator (G) embeds the target data with a fixed watermark. The watermarked data is then combined with the rest of the Diffusion dataset and the Diffusion model is trained as per typical methods. A large number of images are randomly generated using the trained Diffusion model and the resulting images are passed through an image Classifier (C) and watermark Decoder (D).}
\label{fig:Diffusion_Watermarking}
\end{figure}
\section{Method}
\subsection{Watermarking}
In this section, we outline our image watermarker and decoder system. We specifically focus on adding a fixed watermark image to a set of images we wish to protect. Although it is possible to condition the watermarker network on the image, similar to various Deep Steganography methods, it is not necessary for our purposes, as we do not aim to encode a large amount of information in the images. Furthermore, if the process by which the watermark is added to the image is conditioned on the image, the specific augmentation of the image features will differ from image to image. A diffusion model trained on images watermarked in this way would need to learn this potentially complex process. This might become impossible for the diffusion model to achieve if the set of images that are watermarked in the diffusion dataset is small. By generating a single watermark image and adding it to all of the images within a group, we simplify the relationship between images and watermarks. This, in turn, will make it easier for a diffusion model to produce the watermark, as we've created a constant signal for all images within a group.\\
If we assume that the group of images to be watermarked shares some unique common feature or attribute, the addition of a fixed watermark can be thought of as an augmentation of the shared image features. From the perspective of the generative process, the watermark features now become a part of the image group's shared features. Any generative process that aims to accurately produce an image with the original shared feature must now also include the watermark. While simply adding the watermark to the image group is the simplest possible method of augmenting the shared features, validating this process is the first step. By doing so, we provide a foundation for more complicated ways of consistently augmenting the shared features 

\subsubsection{Watermark Generator}
The CNN based Watermark Generator takes in a 1D one-hot coded vector/watermark index and produces a fixed watermark the same shape as the image we wish to protect. Both the image $x_i$ and watermark $w_j$ are scaled between $[-1, 1]$ and combined with a simple linear interpolation controlled by a ``watermark scale" parameter $\lambda$.
\begin{equation} \label{eq:watermark_interp}
    x_{(i, j)} = \lambda w_j + (1 - \lambda) x_i
\end{equation}
This scale parameter controls the intensity/perceptibly of the watermark on the image. To increase the imperceptibility of the embedded watermark the output of the Watermark Generator is blurred with a Gaussian blur to remove high frequency components before being added to the target image. 
\subsubsection{Data Augmentation}
As many image generation piplines contain some sort of data augmentation step we also include a data augmentation step before the watermarked image is passed to the Decoder (Figure \ref{fig:Watermarker}) to ensure the Decoder is robust to these. We use horizontal flips, rotations ($-45$ to $+45$ range), blurring and resize-crops (max 75\% zoom) as these are the most commonly used augmentation techniques used for generative models. We show in experiments that the watermarking system must be specifically trained to be robust to these sorts of augmentations (Table \ref{table:CIFAR_Decoder}).
\subsubsection{Watermark Decoder}
The watermark Decoder is a straight forward ResNet based Softmax classifier \cite{he2016deep}. We use a ResNet18 architecture with the final maxpool layer removed for the CIFAR experiments and an unmodified ResNet34 architecture for all other experiments. The Decoder aims to predict the index of the watermark that was embedded into the input image. An additional loss is constructed to ensure the Decoder produces a flat distribution if the input image does not contain a watermark. This loss helps regularise the Decoder and reduces biases on the output producing a cleaner signal when searching for watermarked images. 
\subsubsection{Training}
Both of the watermark Generator ($G$) and Decoder ($D$) are trained via the Decoder loss, with the additional regularisation loss for the Decoder.\\
For $C$ number of watermarks:
\begin{equation} \label{eq:watermark_interp1}
    d^w_{1:C} = D(G(x_i, w_j, \lambda))\\
\end{equation}
\begin{equation} \label{eq:watermark_interp2}
    d_{1:C} = D(x_i)
\end{equation}
Decoder Loss:
\begin{equation} \label{eq:watermark_interp3}
l_d = -\log \frac{\exp(d^w_j)}{\sum_{c=1}^C \exp(d^w_c)}
\end{equation}
Regularisation Loss:
\begin{equation} \label{eq:watermark_interp4}
l_r = -\sum_{k=1}^C \frac{1}{C} \log \frac{\exp(d_k)}{\sum_{c=1}^C \exp(d_c)}
\end{equation}
Total Loss:
\begin{equation} \label{eq:watermark_interp5}
l_t = l_d + l_r
\end{equation}
As the watermark Generator is trained via the Decoder, problems can occur if the Decoder fails to learn how to detect the watermark embedded in the image early on in training. To combat this we start the watermark scale $\lambda$ at a large value, $\sim 0.5$, and quickly anneal it to the target value over a few epochs and then continue training. During training of the watermark Generator and Decoder, \textbf{random} pairs of images and watermarks are selected. It is only during the training of the diffusion model that a correlation between the watermarks and images is created, at which point the watermark Generator and Decoder parameters are fixed.

\subsection{Diffusion Model}
In all experiments we simply implement the Cold Diffusion de-noising method \cite{bansal2022cold} with no additional changes. We treat the training of the Diffusion model as a ``black box" where we can only change the input data used to train the model.
\begin{figure}[ht]
\centering
\includegraphics[width=0.6\textwidth]{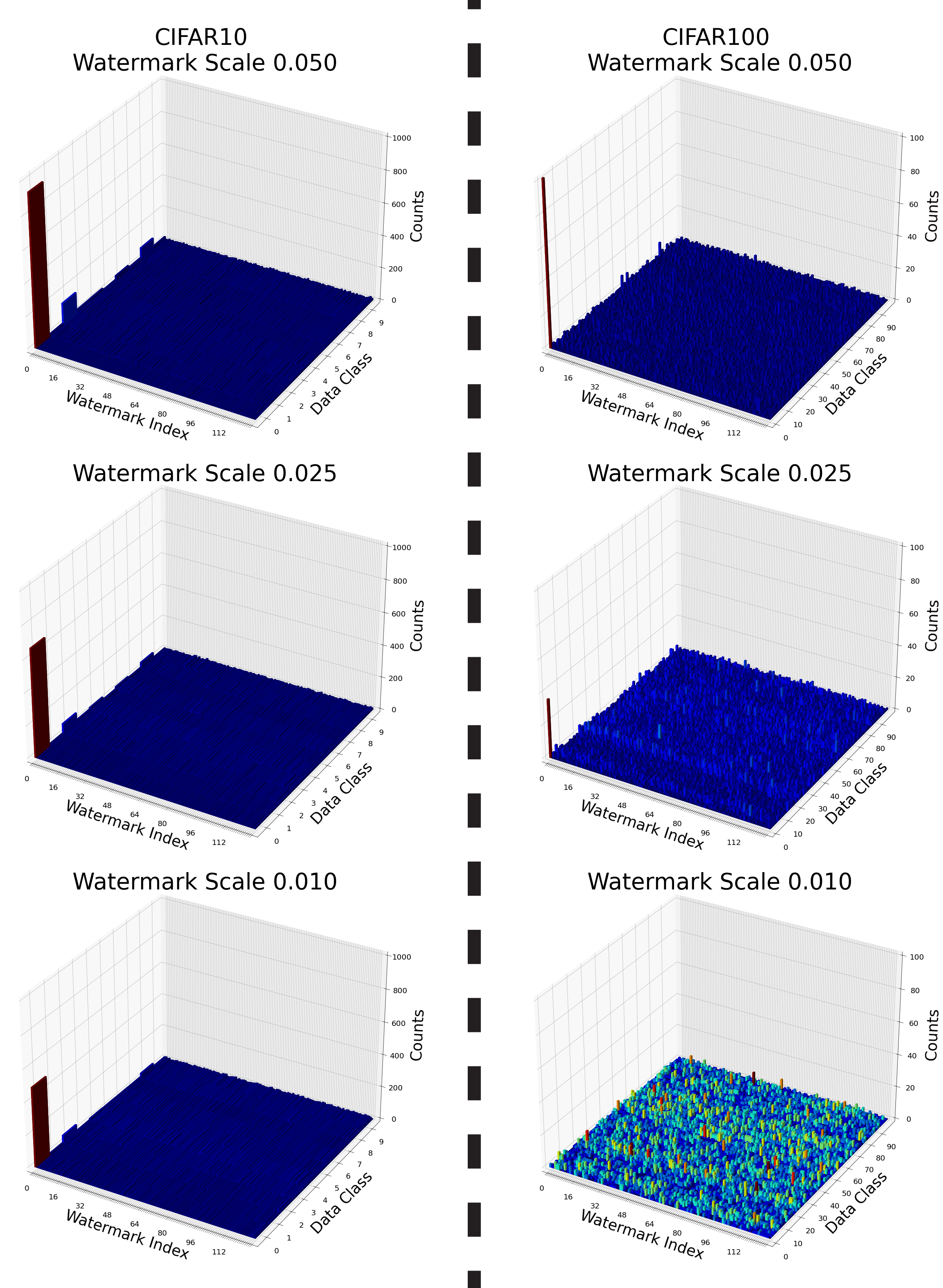}
\caption{2D Histogram of the watermark decoding vs the class predictions for CIFAR10 and CIFAR100 generated images when trained with various watermark scales. For these experiments only images in class 0 were watermarked with watermark index 0.}
\label{fig:CIFAR10_Hists}
\end{figure}
\subsection{Watermarking Diffusion Data}
To generate a correlation between image data and a watermark, we simply pick some aspect(s) of the given dataset and use this to define a sub-group (or multiple subgroups). Each sub-group will be embedded with the same watermark and used to train a diffusion model. This aspect could be high level semantic information; cat, dog, tree, or fine-grained/abstract details like hair colour, or art-style.

\subsection{Chi-Squared Test}
To quantify the results from our experiments, and to determine the effectiveness of our method, we randomly generate images using each diffusion model. Using a trained ResNet image classifier we sort each generated image based on the attributes we wish to track. For example for the CIFAR datasets use the pre-defined class splits. We also use the corresponding watermark Decoder to predict which watermark might be present in the image by randomly sampling from the Softmax distribution. The results are combined into a 2D Histogram of the watermark detections vs the class predictions (Figure \ref{fig:CIFAR10_Hists}). A Chi-Squared test is used to determine whether the distribution of any of the watermarks over the data classes is statistically significant. The expected value for the Chi-Squared test is the mean data class counts for all watermarks. The null hypothesis is therefore that the number and distribution over the dataset classes for each watermark is the same. This allows us to determine whether a particular watermark was present in the training data, even with a biased or inaccurate image data classifier.
\begin{figure}[ht]
\centering
\includegraphics[width=0.8\textwidth]{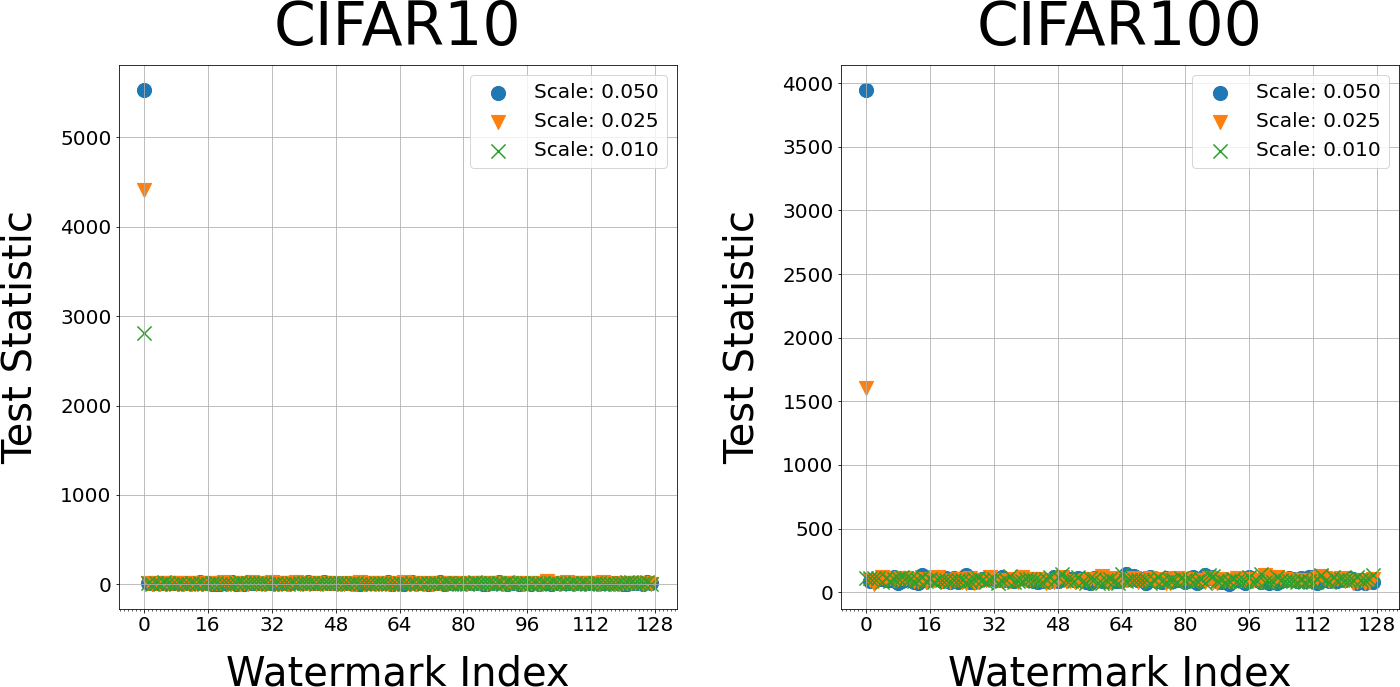}
\caption{The Chi-Squared test statistic vs the watermark index for images generated using the CIFAR10 and CIFAR100 datasets. A larger $\lambda$ value, corresponding to a more obvious watermark, resulting in a stronger signal in the generated images. For these experiments class 0 is embedded with watermark 0.}
\label{fig:CIFAR_TS_index}
\end{figure}

\section{Experiments}
\begin{table}[ht]
\centering
\begin{tabular}{lccccc}
\hline\noalign{\smallskip}
 Dataset & Aug Trained & $\lambda$ & WM MSE ($10^{-5}$) & No-Aug Acc & Aug Acc \\
\noalign{\smallskip}
\hline
\noalign{\smallskip}
CIFAR10 & Y & 0.050 & 209 & 1.00 & 0.89\\
CIFAR10 & Y & 0.025 & 51 & 0.99 & 0.85\\
CIFAR10 & Y & 0.010 & 8.4 & 0.68 & 0.50\\
\hline
CIFAR10 & N & 0.025 & 48 & 1.00 & 0.49\\
\hline
\hline
CIFAR100 & Y & 0.050 & 210 & 1.00 & 0.88\\
CIFAR100 & Y & 0.025 & 53 & 0.98 & 0.83\\
CIFAR100 & Y & 0.010 & 8.6 & 0.74 & 0.56\\
\hline
CIFAR100 & N & 0.025 & 49 & 0.99 & 0.47\\
\hline
\end{tabular}
\caption{
Quantitative results of Watermark Decoder Accuracy on test-set. All models are trained to generate and decode 128 watermarks. The results show the the impact of changing the watermark scale $\lambda$ and the effect of augmenting the watermarked images during training.
}
\label{table:CIFAR_Decoder}
\end{table}
\setlength{\tabcolsep}{1.4pt}

\setlength{\tabcolsep}{4pt}
\begin{table}[ht]
\centering
\begin{tabular}{lccc}
\hline\noalign{\smallskip}
 Dataset & $\lambda$ & p-value $\downarrow$\\
\noalign{\smallskip}
\hline
\noalign{\smallskip}
CIFAR10 & 0.050 & $<<10^{-6}$\\
CIFAR10 & 0.025 & $<<10^{-6}$\\
CIFAR10 & 0.010 & $<<10e^{-6}$\\
\hline
CIFAR100 & 0.050 & $<<10^{-6}$\\
CIFAR100 & 0.025 & $<<10^{-6}$\\
CIFAR100 & 0.010 & $0.194$\\
\hline
CIFAR10 + Aug & 0.025 & $<<10^{-6}$\\
CIFAR10 Batch 1 & 0.025 & $5.38\times10^{-5}$\\
\hline
\end{tabular}
\caption{
Quantitative results of the signal strength of a single watermark for various $\lambda$ values. For each experiment 12800 images were generated and the watermark index and data class were predicted. Chi-Squared test statistics and p-values  of the target watermark index are given.
}
\label{table:CIFAR_Diffusion}
\end{table}
\setlength{\tabcolsep}{1.4pt}

\subsection{Watermark Strength}
\subsubsection{Decoder Accuracy}
To test the sensitivity of our watermarking system, we train a number of watermark Generators and Decoders with various watermarking scales $\lambda$ on the CIFAR10 and CIFAR100 datasets \cite{krizhevsky2009learning}. We also test the impact of adding a data augmentation step after watermarking. To create the test results we perform 5 epochs over the corresponding test-sets, choosing random image/watermark pairs each time as well as a different data augmentation method. For each experiment 128 watermarks are used.\\
Table \ref{table:CIFAR_Decoder} shows that for both datasets the Decoder is able to accurately decode the correct watermark. This is the case even when data augmentation is applied, so long as the system was trained with data augmentation. The watermarks embedded in the images remain imperceptible for most images (Figure \ref{fig:wm_imgs}) and only begins to become discernible when $\lambda=0.050$. We also note the effects of data augmentation on the watermarks themselves. When training with data augmentations, the watermarks become more unique and symmetrical.

\subsubsection{Generated Signal Strength}
To determine the survivability of the watermarked images during the diffusion process, we train several diffusion models with different $\lambda$ values. The data for each diffusion model is watermarked using one of the previously trained CIFAR10/100 watermark generators. For each experiment a single class (class index 0) of the dataset is encoded with the same watermark. As CIFAR10 and CIFAR100 datasets are equally split into 10 and 100 classes, we can also observe the effect of watermarking different percentages of the whole dataset.
It is clear from the results of these experiments that the generated images of the target class overwhelmingly contain the target watermark (Table \ref{table:CIFAR_Diffusion}, Figure \ref{fig:CIFAR_TS_index}). Importantly, results also show that no other watermark index shares any strong correlation to a specific data class (Figure \ref{fig:CIFAR10_Hists}). Increased counts of the target watermark in other data classes are likely to be miss-classifications by the image classifier. For example, for the CIAFAR10 dataset the target data class, ``airplane", is visually similar to the classes of the other two main peaks ``bird" and ``ship". A discernible signal is even present in the CIFAR100 experiments, were only 1\% of the total dataset is watermarked. The p-value only becomes insignificant when $\lambda = 0.01$. 
From these experiments we determine that a watermark scale of $\lambda=0.025$ provides a good trade-off between imperceptibility and signal strength and is used for all subsequent experiments.
\begin{figure}
     \centering
     \begin{subfigure}[ht]{0.45\textwidth}
         \centering
         \includegraphics[width=\textwidth]{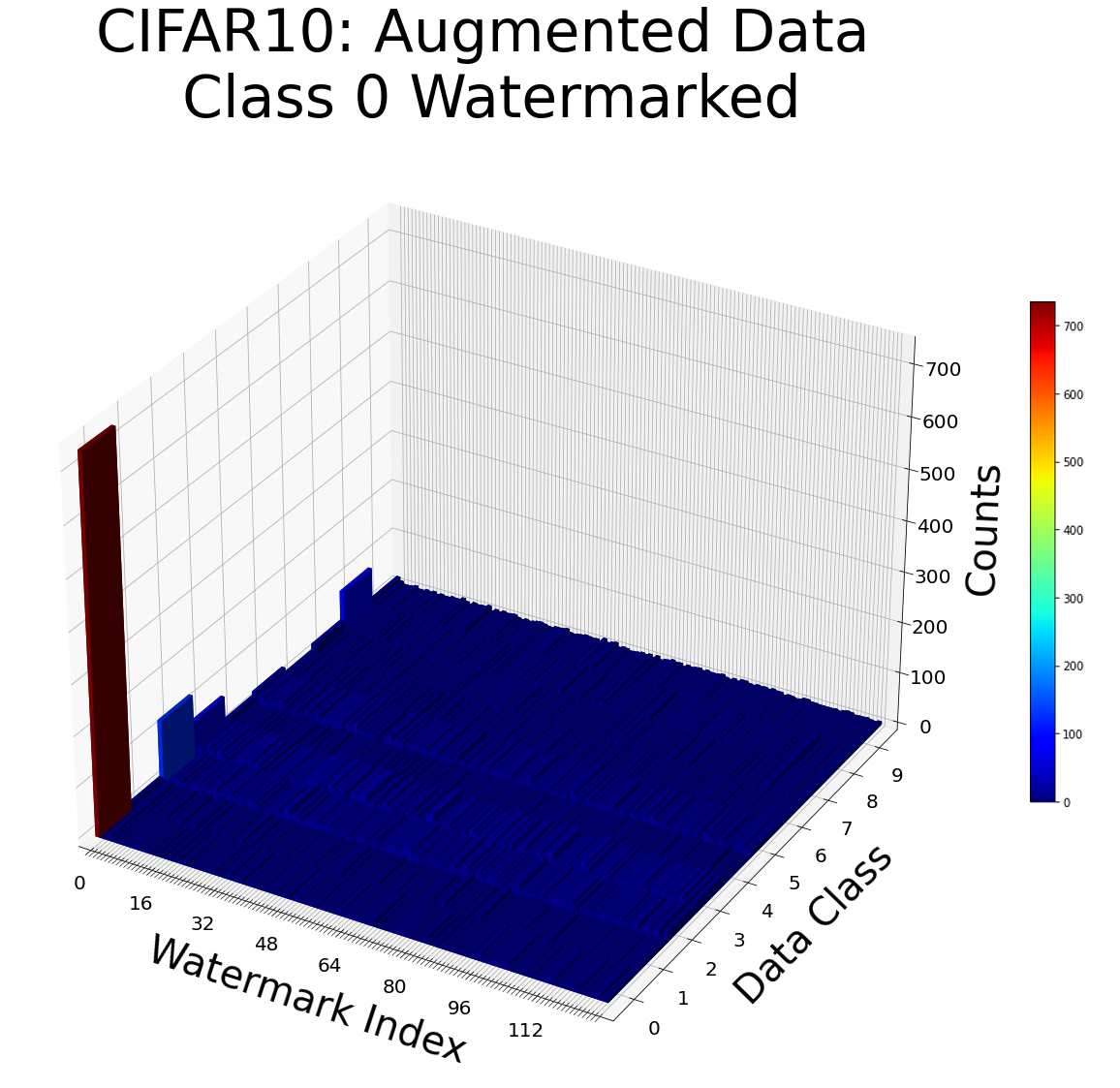}
         \caption{}
         \label{fig:CIFAR10_Aug_025_1c_hist}
     \end{subfigure}
     \begin{subfigure}[ht]{0.45\textwidth}
         \centering
         \includegraphics[width=\textwidth]{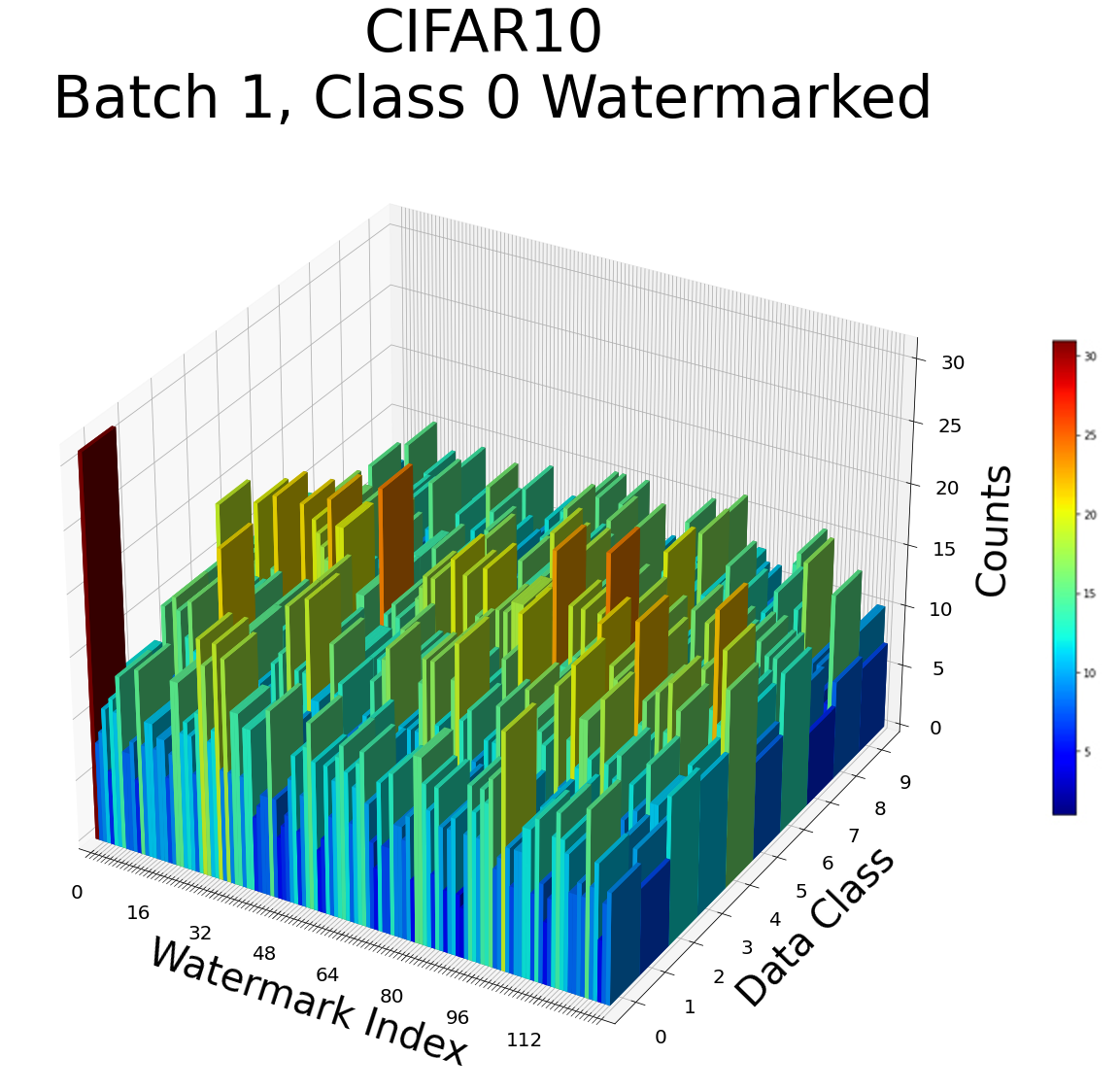}
         \caption{}
         \label{fig:CIFAR10_025_1c_batch1_hist}
     \end{subfigure}
        \caption{a) Class 0 of CIFAR10 is embedded with watermark 0 and all data is heavily augmented before being used to train a diffusion model. b) Only Class 0 images of Batch 1 ($2\%$ of total data) of CIFAR10 are watermarked prior to training the diffusion model.}
        \label{fig:Special_Cases}
\end{figure}

\subsubsection{Diffusion with Augmented Watermarks}
To simulate pre-processing steps when training a generative model we perform data augmentation to the watermarked images before they are used to train the diffusion model. We use a watermark Generator/Decoder that was trained with augmentation to embed the target images with the watermark. The data augmentations used are the same type used in training the watermark Generator and Decoder and are randomly applied every epoch. We mark all of class 0 of the CIFAR10 dataset with watermark index 0, the same as previous experiments. Even though the data augmentations causes a change in the appearance of watermarks, results show that the Diffusion model is able to generate them accurately enough for the Decoder to identify them (Figure \ref{fig:CIFAR10_Aug_025_1c_hist}).
\subsubsection{Partial Class Watermarking}
In a real-world scenario, it is likely that only some of the images that share a particular aspect within a dataset are watermarked To simulate this situation we experiment with only embedding a watermark into a fraction of images within a CIFAR10 class. We utilise the the pre-defined ``data batches" for CIFAR10 to perform the split and only watermark images of class 0 within batch 1. As CIFAR10 is split into 5 relatively equal ``data batches", we are watermarking $20\%$ of class 0, accounting for $2\%$ of the overall dataset. Even in this case a detectable signal is present  with $p=5.38\times10^{-5}$ (Table \ref{table:CIFAR_Diffusion}). However it is important to note that the previously strong signal is greatly diminished, with the watermark/class peak being less obvious than the CIFAR100 experiments which had a similar percentage of images watermarked (Figure \ref{fig:CIFAR10_025_1c_batch1_hist}). This indicates that image watermarking is less effective when watermarked images share same features as many un-watermarked images within the same dataset.

\begin{figure}[ht]
\centering
\includegraphics[width=0.9\textwidth]{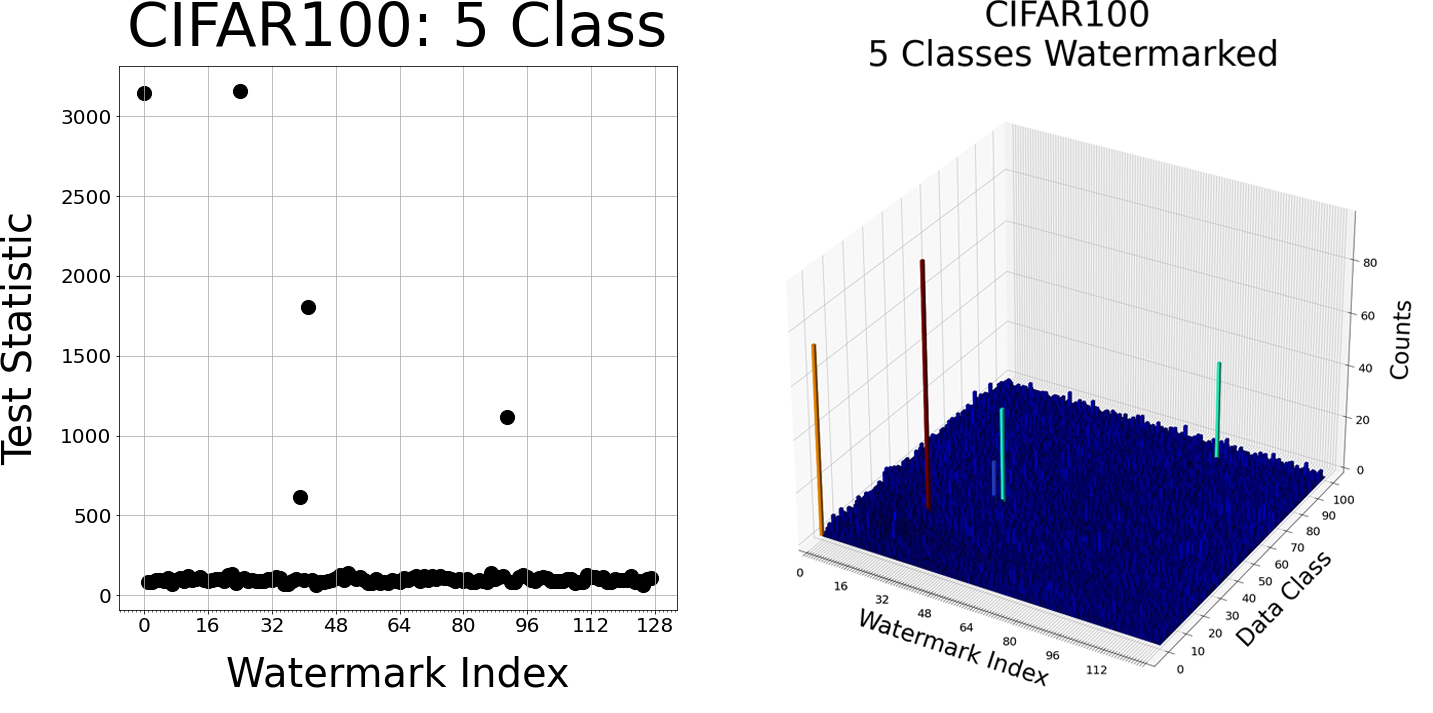}
\caption{Diffusion results using the CIFAR100 dataset with 5 classes embedded with a unique watermark. Left: Chi-Squared test statistic for each watermark index. Right: 2D Histogram of Watermarks vs the predicted data class.}
\label{fig:CIFAR100_5C_plots}
\end{figure}

\subsection{Multiple Watermarks Per Dataset}
\begin{figure}[ht]
\centering
\includegraphics[width=0.9\textwidth]{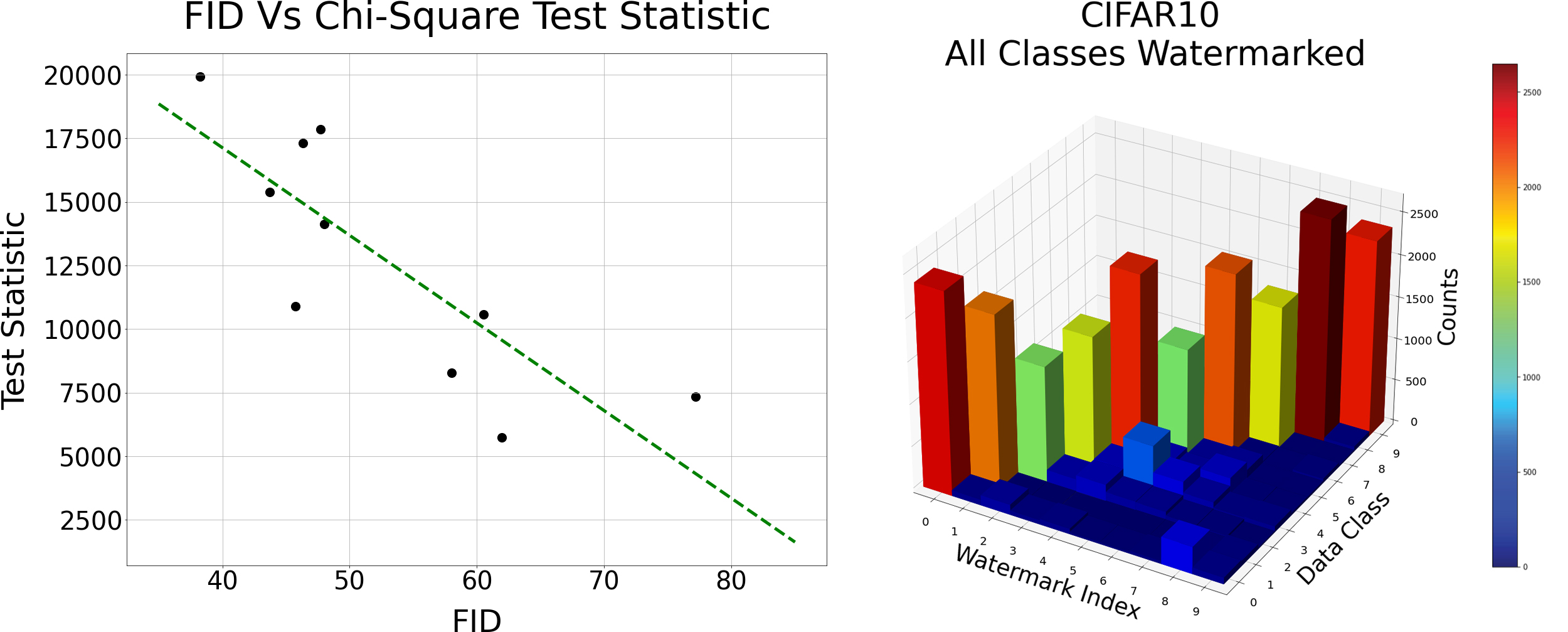}
\caption{Diffusion results using the CIFAR10 dataset with each class embedded with a unique watermark. Left: Chi-Squared test statistic vs the FID score of the corresponding data Class. Right: 2D Histogram of the first 10 Watermarks vs the predicted data class.}
\label{fig:CIFAR10_AllC_plots}
\end{figure}

To further demonstrate how the diffusion model maintains the correlation between the watermarks and data features, we undertake experiments utilising multiple watermarks. In this situation, multiple groups within the dataset are watermarked, with each group embedded with a unique watermark. We start with the simple case of marking each class of CIFAR10 with a unique watermark (10\% of data per-watermark). We also perform the more realistic experiment of watermarking only 5 classes from CIRFAR100 (1\% of data per-watermark). For convenience in both of these experiments the data class index is also used as the watermark index. Results from both experiments show a good correlation between the target watermarks and the corresponding image data class (Figure \ref{fig:CIFAR10_AllC_plots} and \ref{fig:CIFAR100_5C_plots}). 
\subsubsection{FID Vs Signal Strength}
From the 2D Histograms of this experiment it is evidant that some class/watermark pairs have a stronger signal strength then others (Figure \ref{fig:CIFAR100_5C_plots}). This seems to be due to the difference in generated image quality between classes, potentially due to different levels of image complexity. To investigate we plot the per-class FID score \cite{heusel2017gans} against the Chi-Squared test statistic for the corresponding watermark index. The original \textbf{Unwatermarked} images are used as reference to compute the FID score, with each FID score computed independantly for each class. These results shows a clear relationship between image quality and the watermarks signal strength (Figure \ref{fig:CIFAR10_AllC_plots}). This suggests that a higher quality generative model, that is able to better capture the data-distribution, will not only produce higher quality images, but a stronger watermark signal.
\begin{figure}[ht]
\centering
\includegraphics[width=0.9\textwidth]{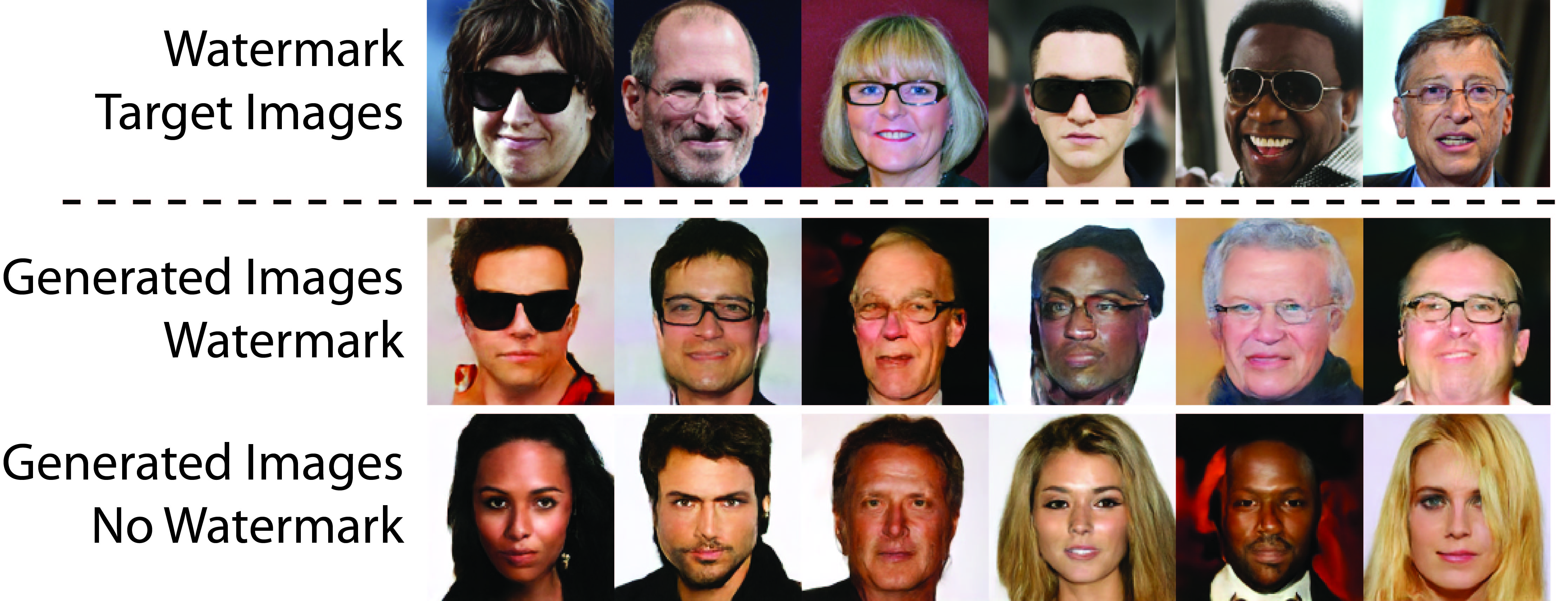}
\caption{Only CelebAHQ images with the ``Eyeglasses" attribute are watermarked before being used to train a Diffusion model. Using a watermark Decoder we can pick out generated images that have a high confidence of having the target watermark.}
\label{fig:CelebA_glasses_target}
\end{figure}
\subsection{Fine-grained Attributes}
\begin{figure}[ht]
\centering
\includegraphics[width=0.8\textwidth]{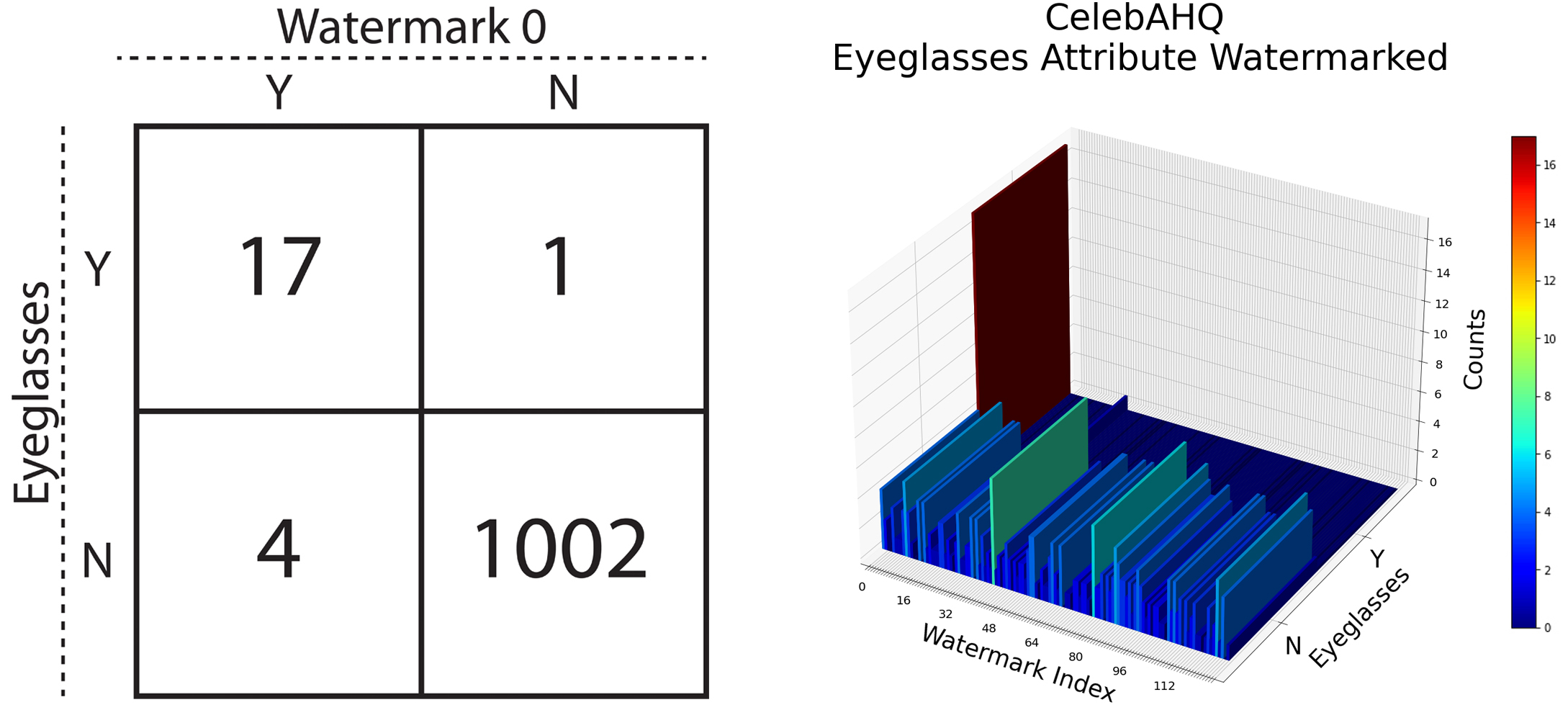}
\caption{Diffusion results using the CelebAHQ dataset with all images with the ``Eyeglasses" attribute ($4.8\%$ of data) embedded with watermark 0. Left: 2x2 contingency table between the detections of Eyeglasses and watermark 0 in the CelebAHQ generated images. Right: 2D Histogram of Watermarks vs Eyeglasses detected in the CelebAHQ generated images.}
\label{fig:CelebA_Eyeglasses_Plots}
\end{figure}

The CIFAR experiments allow us to watermark groups of images that share prominent features, we now investigate correlating a watermark with fine-grain image information. To do so we train a watermark Generator and Decoder to embed 512 watermarks using the CelebAHQ dataset at a 128x128 resolution \cite{CelebAMask-HQ}. Once trained we embed all images of the dataset with the ``Eyeglasses" attribute (accounting for $4.8\%$ of all data) with watermark 0. A Diffusion model is then trained on these 128x128 resolution images and once training is complete 1024 images are randomly generated. To sort the generated images by the target attribute a ResNet18 classifier is trained to detect eyeglasses in the CelebAHQ images. All but one of the generated images with eyeglasses detected were decoded as having watermark 0 \ref{fig:CelebA_Eyeglasses_Plots}. To quantify this result we simplify the 2D Histogram into a 2x2 contingency table and perform a Fisher's exact test (Figure \ref{fig:CelebA_Eyeglasses_Plots}). The calculated p-value $2.9\times10^{-32}$ signifies that there is a clear relationship between images with eyeglasses and watermark 0 (Figure \ref{fig:CelebA_glasses_target}). 

\section{Conclusion}
In this paper, we introduce the idea of consistently and imperceptibly watermarking images to track their use in training generative diffusion models. Our method specifically targets the creators of digital content and offers a way to track the use of copyrighted material in artificially generated content. We hope that this work encourages future research on the non-consensual use of digital content. We believe future work should focus on the current problem areas, such as using less watermarked data to train diffusion models, as well as developing new methods to embed watermarks and detect the signal in generated outputs. The use of improved methods for constantly augmenting the shared features of the image group may allow for more aggressive watermarking. Such methods could still remain imperceptible but apply more complicated embedding techniques. For example, generative techniques could subtly imprint a consistent pattern onto the skin tone of pictures of an individual's face. The exact pixel changes would differ per image, but the exact pattern would remain the same (light and dark regions in the same area of the face). We hope that in the future, such a method may be employed to help manage the use of digital content online and provide proper attribution to creators.

\bibliographystyle{alpha}
\bibliography{ref}

\newpage
\begin{appendices}
\section{Additional Results}

\subsection{CelebA HQ}
\begin{figure}[ht]
\centering
\includegraphics[width=0.90\textwidth]{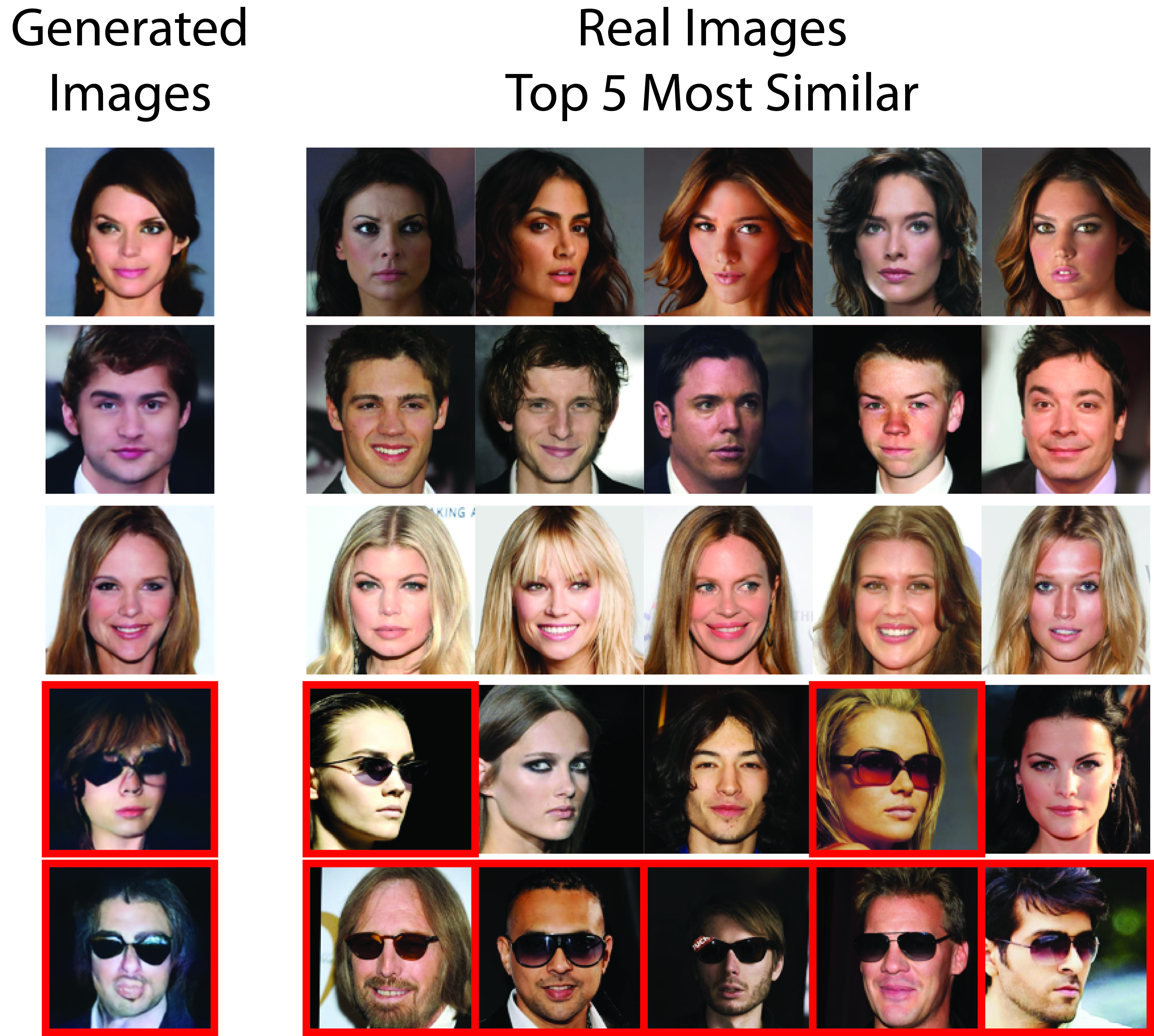}
\caption{Generated images with the top 5 closest matches. The features from a pre-trained ResNet34 were used to determine the distance from each of the generated images to all of the training images. The red border indicates images where the target watermark was decoded (generated images) or added (real images).}
\label{fig:CelebAHQ_Top5}
\end{figure}

Figure \ref{fig:CelebAHQ_Top5} shows that even when the Diffusion model is creating novel images, the target watermarks are present and correlated to the features of the target images.

\begin{figure}
\centering
\includegraphics[width=0.90\textwidth]{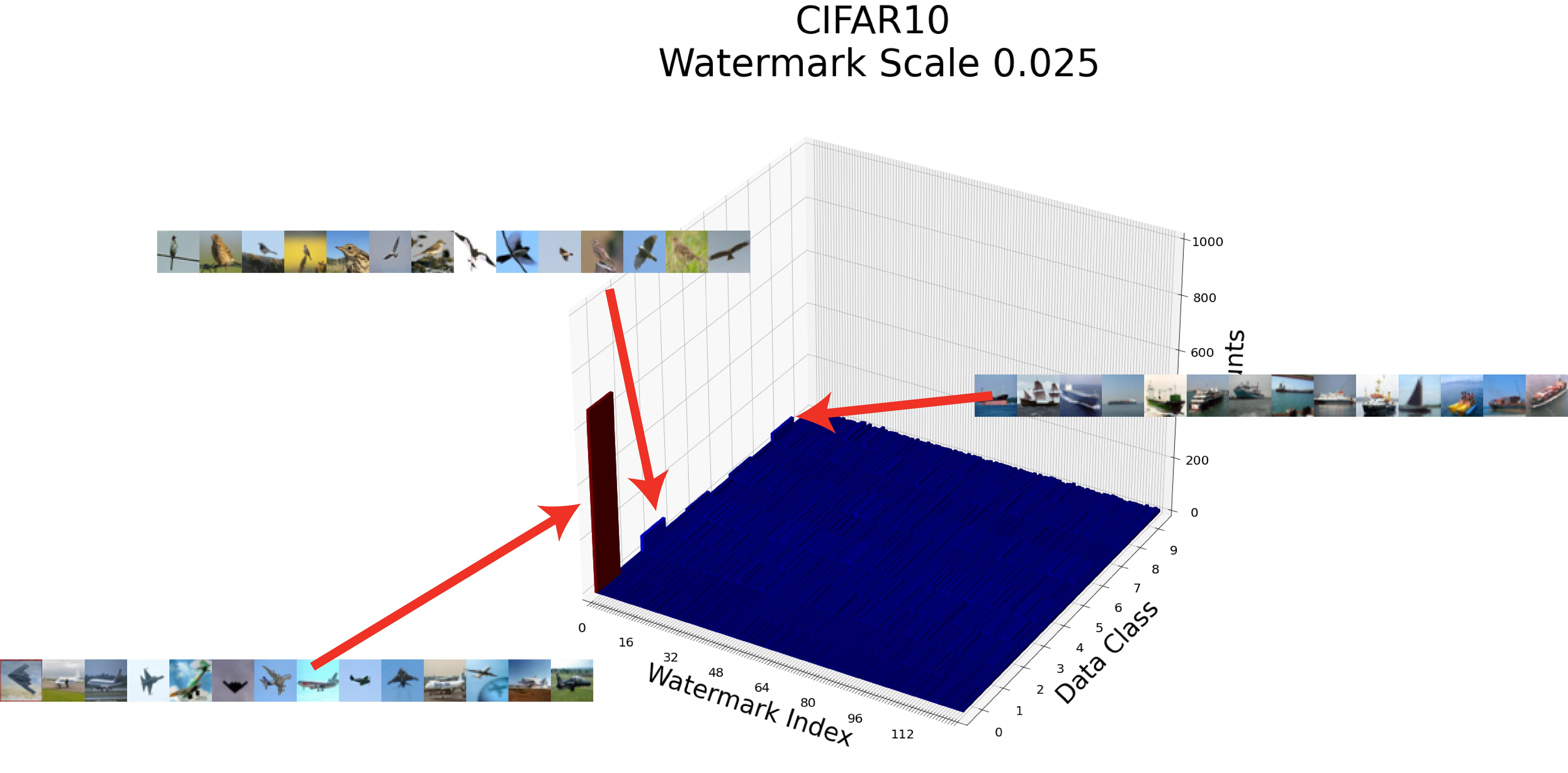}
\caption{CIFAR10 Histogram, showing that the additional peaks are from visually similar classes.}
\label{fig:CIFAR10_Similar_Classes}
\end{figure}

\begin{figure}
     \centering
     \begin{subfigure}[ht]{0.80\textwidth}
         \centering
         \includegraphics[width=\textwidth]{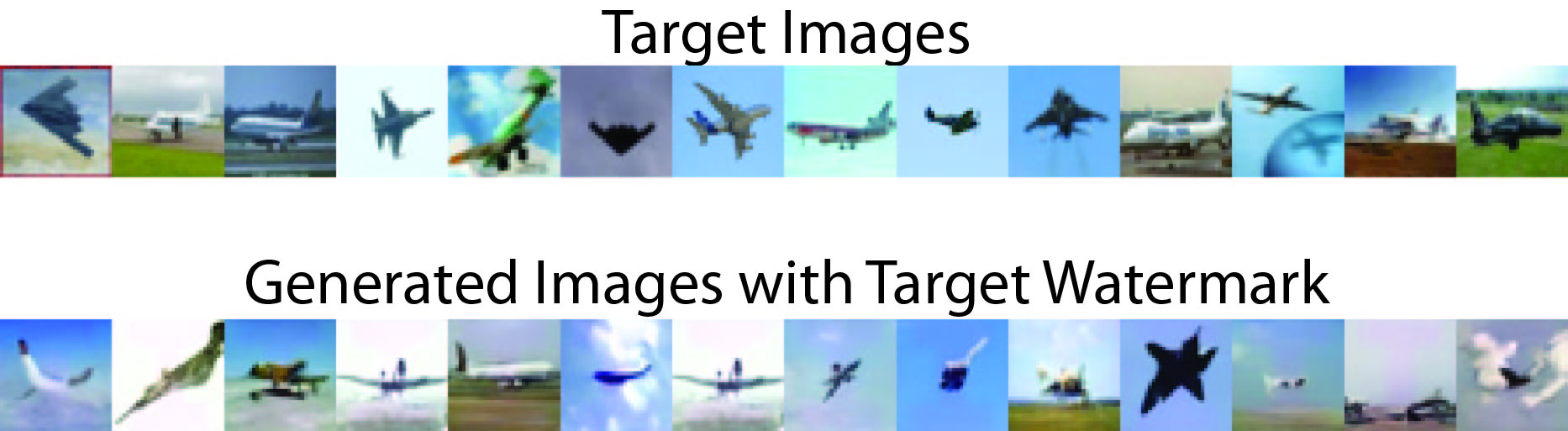}
         \caption{}
         \label{fig:CIFAR10_Class0}
     \end{subfigure}
     \begin{subfigure}[ht]{0.80\textwidth}
         \centering
         \includegraphics[width=\textwidth]{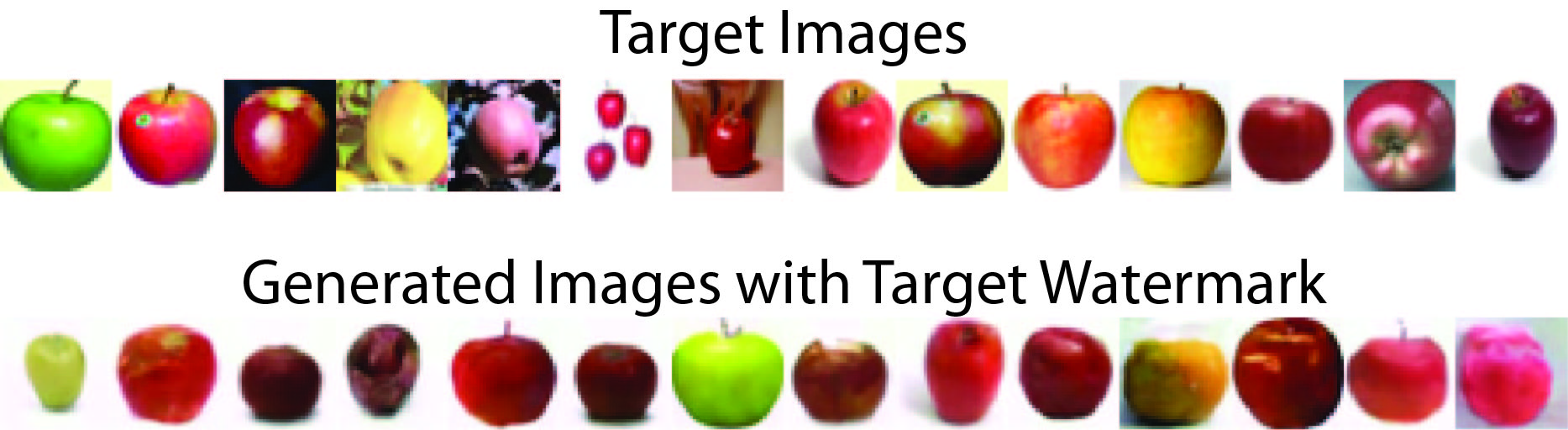}
         \caption{}
         \label{fig:CIFAR100_Class0}
     \end{subfigure}
        \caption{CIFAR10 (a) and CIFAR100 (b) target images as well as generated images that decode to the target watermark. $\lambda = 0.025$ in both cases.}
        \label{fig:Special_Cases2}
\end{figure}

\subsection{CIFAR}
Additional results showing the target images to be watermarked and the generated images that the watermark Decoder decoded to the target watermark with a high confidence (Figure \ref{fig:CIFAR10_Class0}, \ref{fig:CIFAR100_Class0}). Figure \ref{fig:CIFAR10_Similar_Classes} shows that the additional peaks in the CIFAR10 histogram most likely come from mis-classifications of the image data.

\begin{figure}
     \centering
     \begin{subfigure}{0.45\textwidth}
         \centering
         \includegraphics[width=\textwidth]{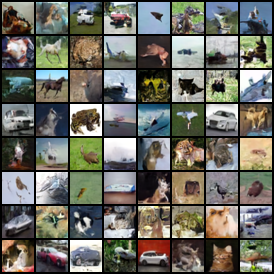}
         \caption{}
         \label{fig:CIFAR10_Diffusion}
     \end{subfigure}
     \begin{subfigure}{0.45\textwidth}
         \centering
         \includegraphics[width=\textwidth]{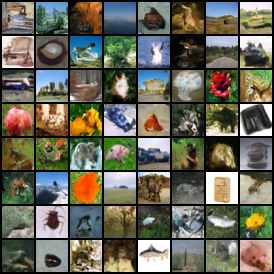}
         \caption{}
         \label{fig:CIFAR100_Diffusion}
     \end{subfigure}
        \caption{Generated CIFAR10 (a) and CIFAR100 (b) images.}
        \label{fig:Diffusion}
\end{figure}

\section{Network Architecture}
\subsection{Watermark Generator}
The watermark Generator is a Res-CNN that takes in a watermark index and outputs a unique watermark the same size as the target image.
\begin{figure}[ht]
\centering
\includegraphics[width=\textwidth]{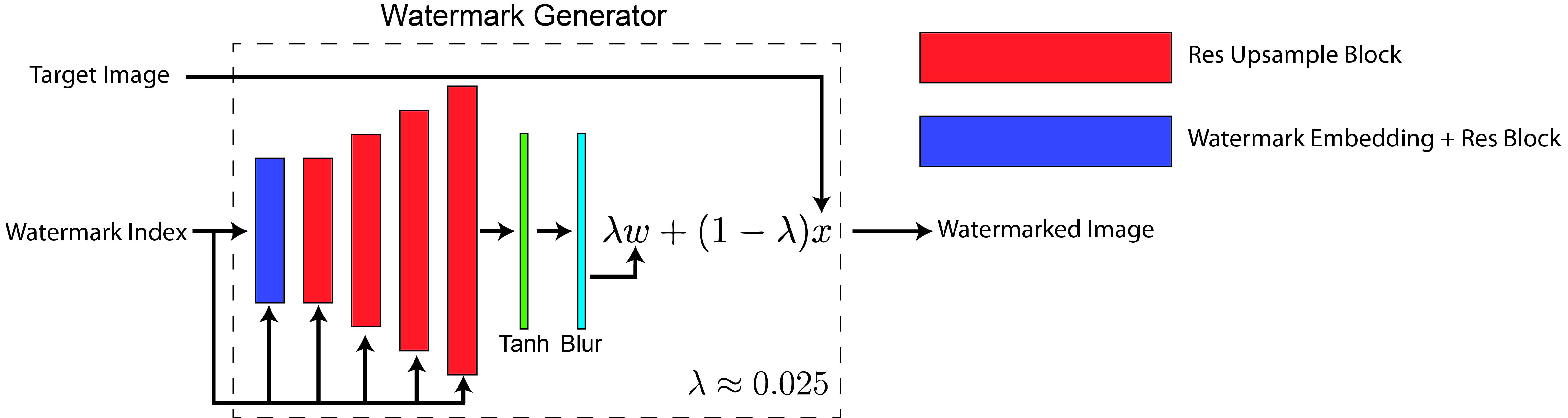}
\caption{The watermark Generator is a CNN made up of Residual layers that up-sample the given Watermark embedding. The output is passed through a Tanh layer so the values are bound between (-1, 1), like the image data. This allows us to easily combine the watermark with the image and ensures that the result is also bounded between (-1, 1) without clipping. Gaussian Blurring is applied to remove high frequency components of the watermark, making the watermark more imperceptible.}
\label{fig:Watermark_Generator_details}
\end{figure}\\

\begin{figure}
     \centering
     \begin{subfigure}[ht]{0.49\textwidth}
         \centering
         \includegraphics[width=\textwidth]{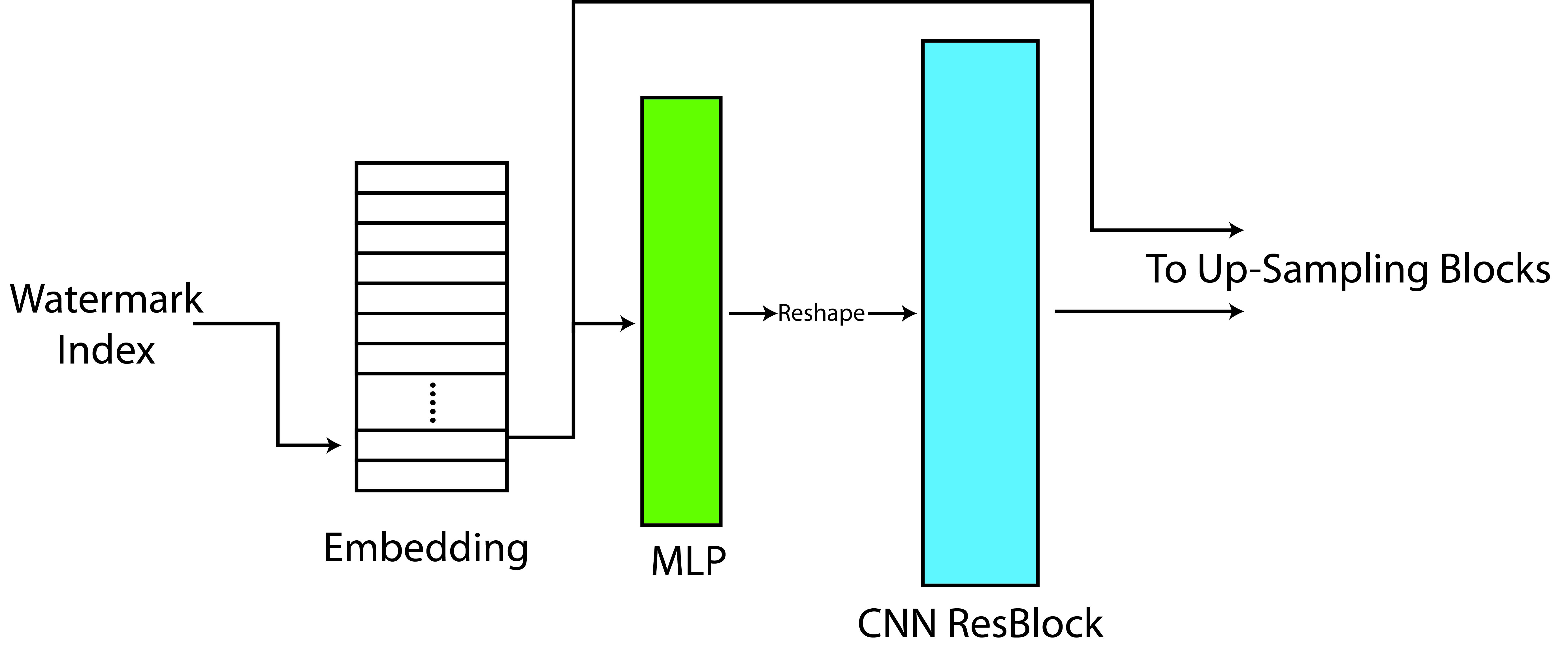}
         \caption{}
         \label{fig:Embedding_ResBlock}
     \end{subfigure}
     \begin{subfigure}[ht]{0.49\textwidth}
         \centering
         \includegraphics[width=\textwidth]{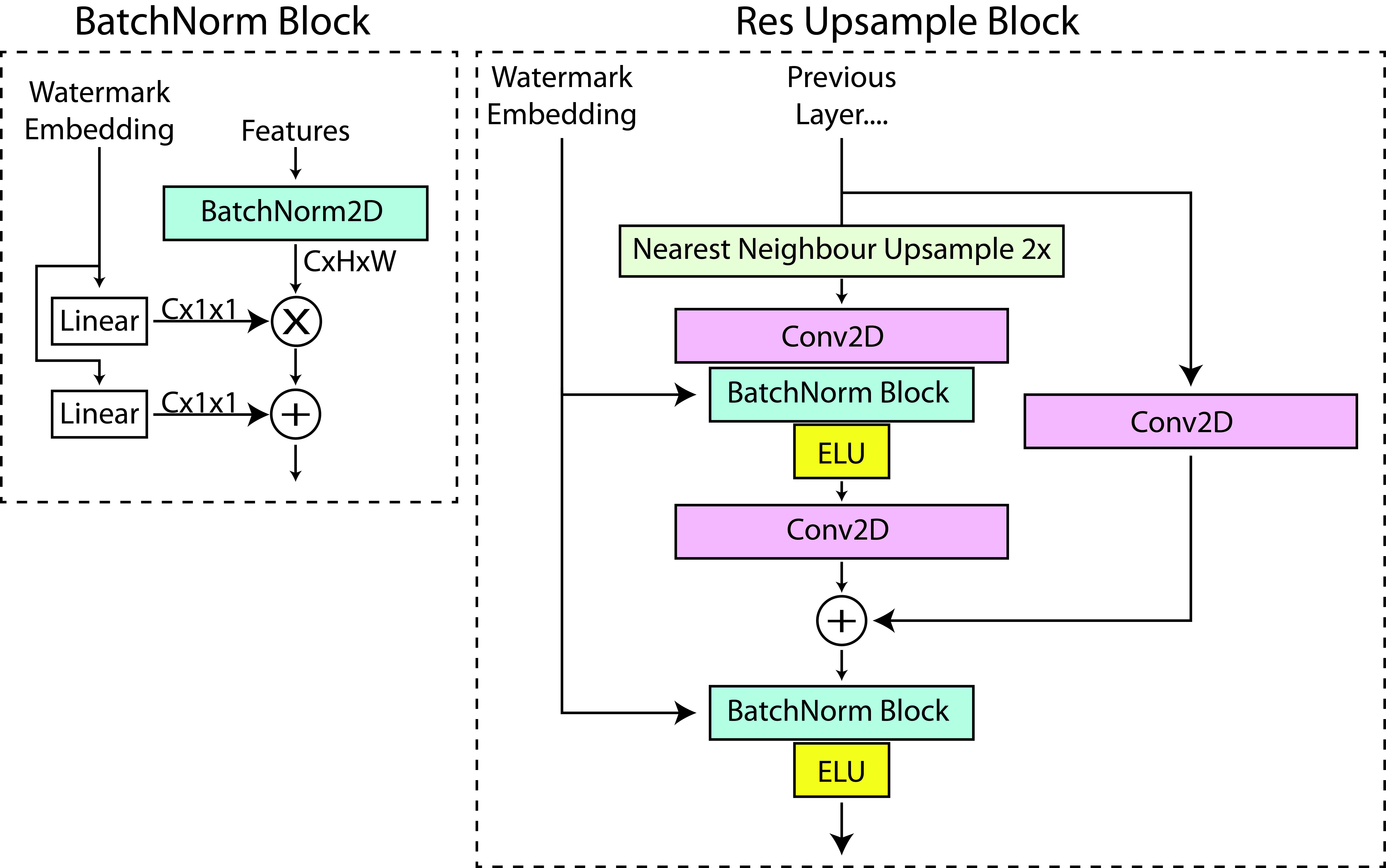}
         \caption{}
         \label{fig:UpSampling_Block}
     \end{subfigure}
        \caption{a) The embedding corresponding to the watermark index is passed through a Multi-layer Perceptron (MLP) and the output is reshaped and passed through a CNN ResBlock. The CNN ResBlock has the same layout as the Res Up-sampling block, without the Upsampling. b) The architecture of the Res Up-Sampling block. The watermark embedding is used to augment the scale and bias of the outputs from the BatchNorm layers.}
        \label{fig:Blocks}
\end{figure}
\section{Experiment Details}
All models are trained using a single NVIDIA RTX 2080Ti or 3060Ti GPU. All training code is implemented using Pytorch.
\subsection{Watermark Generator and Decoder}
For the CIFAR10 and CIFAR100 experiments a ResNet18 architecture with the final maxpool layer removed is used as the watermark Decoder. For the CelebA experiments an unmodified ResNet34 architecture is used. The value of $\lambda$ is initialised to a large value (0.5) and is reduce early in training (over about 30 epochs) to the the final value. The exact values of $\lambda$ used are specified in the experiments section, but a typical value of $0.025$ works well.\\
\begin{table}[ht]
\centering
\begin{tabular}{l|c|c|c|c}
\hline\noalign{\smallskip}
 Watermarks & Blur Kernel Size/Sigma &  Epochs & G Channel Scale & G Block Widths\\
\noalign{\smallskip}
\hline
\noalign{\smallskip}
128 & $k=5\times5$, $\sigma=2$ & 200 & 64 & (4, 2, 1)\\
\end{tabular}
\caption{
CIFAR10 and CIFAR100 hyperparameters for training the watermark Generator (G) and Decoder (D). The block widths correspond to the width of the input features to each of the Res Up-sampling blocks within the watermark Generator.}\label{table:cifar_watermark}
\end{table}

\begin{table}[ht]
\centering
\begin{tabular}{l|c|c|c|c}
\hline\noalign{\smallskip}
 Watermarks & Blur Kernel Size/Sigma &  Epochs & G Channel Scale & G Block Widths\\
\noalign{\smallskip}
\hline
\noalign{\smallskip}
512 & $k=11\times11$, $\sigma=4$ & 200 & 64 & (4, 2, 2, 2, 1)\\
\end{tabular}
\caption{
CelebA hyperparameters for training the watermark Generator (G) and Decoder (D). The block widths correspond to the width of the input features to each of the Res Up-sampling blocks within the watermark Generator.}\label{table:celebwatermark}
\end{table}

\subsection{Diffusion Models}
The Diffusion models were trained via De-Noising Cold Diffusion and a standard Diffusion U-Net model was used. Training parameters are in Table \ref{table:cifar_diffusion} and \ref{table:celeba_diffusion}. 

\begin{table}[ht]
\centering
\begin{tabular}{l|c|c|c|c|c}
\hline\noalign{\smallskip}
 Epochs & Batch Size & Channel Scale &  Block Widths & Learning Rate & Diffusion Steps\\
\noalign{\smallskip}
\hline
\noalign{\smallskip}
900 & 64 & 128 & (1 2 2 4) & $2e^{-5}$ & 100\\
\end{tabular}
\caption{
Hyperparameters for training Diffusion Model for each of the CIFAR10 and CIFAR100 experiments.}\label{table:cifar_diffusion}
\end{table}

\begin{table}[ht]
\centering
\begin{tabular}{l|c|c|c|c|c}
\hline\noalign{\smallskip}
 Epochs & Batch Size & Channel Scale &  Block Widths & Learning Rate & Diffusion Steps\\
\noalign{\smallskip}
\hline
\noalign{\smallskip}
250 & 16x2 & 64 & (1 2 2 4 4 8) & $2e^{-5}$ & 100\\
\end{tabular}
\caption{
Diffusion Model hyperparameters for the 128x128 CelebA-HQ experiment. Gradient accumulation is used to achieve a batch size of 32 over 2 forward passes.}\label{table:celeba_diffusion}
\end{table}

\end{appendices}
\end{document}